\documentclass[aip,amsmath,amssymb,reprint,groupedaddress]{revtex4-1}
\pdfoutput=1     

\usepackage{graphicx}
\usepackage{siunitx} 
\usepackage{array} 
\usepackage{xcolor}

\graphicspath{ {figures/} }

\newcommand{\kB}{k_{\mathrm{B}}}
\newcommand{\kT}{\kB T}

\newcommand{\vek}[1]{\boldsymbol{#1}}          

\newcommand{\ffrac}[2]{\frac{\displaystyle #1}{\displaystyle #2}}

\begin{document}

\title{Mechanisms of DNA Hybridization: Transition Path Analysis of a Simulation-Informed Markov Model}
\author{Raymond Jin}
\author{Lutz Maibaum}
\email{maibaum@uw.edu}
\affiliation{Department of Chemistry, University of Washington, Seattle, WA 98195, USA}
\begin{abstract}
Complementary DNA strands in solution reliably hybridize to form stable duplexes. We study the kinetics of the hybridization process and the mechanisms by which two initially isolated strands come together to form a stable double helix. We adopt a multi-step computational approach. First, we perform a large number of Brownian dynamics simulations of the hybridization process using the coarse-grained oxDNA2 model. Second, we use these simulations to construct a Markov State Model of DNA dynamics that uses a state decomposition based on the inter-strand hydrogen bonding pattern. Third, we take advantage of Transition Path Theory to obtain quantitative information about the thermodynamic and dynamic properties of the hybridization process. We find that while there is a large ensemble of possible hybridization pathways there is a single dominant mechanism in which an initial base pair forms close to either end of the nascent double helix, and the remaining bases pair sequentially in a zipper-like fashion. We also show that the number of formed base pairs by itself is insufficient to describe the transition state of the hybridization process.
\end{abstract}

\maketitle

\section*{\label{sec:intro}Introduction}
Deoxyribonucleic acid (DNA) is a biopolymer that contains the information of life. Its crucial role in biology and its many uses in technological applications such as DNA origami~\cite{Seeman1982,Rothemund2006} and nanoparticle linker~\cite{Mirkin1996,Chou2014} depend on DNA's ability to hybridize: two single-stranded DNA (ssDNA) molecules bind to form a double-stranded DNA (dsDNA) duplex. The duplex is stabilized by hydrogen bonds between the nucleobases adenine (A), thymine (T), guanine (G), and cytosine (C).

The mechanism of hybridization remains a topic of current research.
Classically, hybridization is thought of as a two-step process in which the strands first form an initial nucleus of hydrogen-bonded base pairs, which is then followed by the strands ``zipping up'' to form the fully hybridized double helix~\cite{Wetmur1968}.
It was later found that at least three base pairs were necessary for a thermodynamically stable binding nucleus~\cite{Porschke1971}.
Recent experiments on RNA, on the other hand, found a threshold of seven base pairs~\cite{Cisse2012}.
A more complex mechanism based on a three-step process has been proposed~\cite{Niranjani2016}. Here the two strands initially bind by non-specific interactions. The first stable base pair is then formed either through a one-dimensional sliding motion in which the strands search for an initial nucleation point, or through internal displacements in which the strands traverse each other in an inchworm-like motion. Once an initial nucleation point is found, the bases zip up to fully hybridize as in the classical mechanism.

It is challenging to observe hybridization in a base-by-base manner in experiments.
Previous work has been done on detection of hybridization in bulk solution using a variety of techniques, including absorption spectroscopy~\cite{Wetmur1968,Parkhurst1995,Wang2014}, surface-enhanced Raman spectroscopy~\cite{Barhoumi2010}, and electrochemistry~\cite{Drummond2003}.
Recently, there have been advances in single molecule experiments to study hybridization using electrochemical techniques~\cite{Sorgenfrei2011} and force probes~\cite{Liphardt2001,Woodside2006}.
There have also been studies of DNA at the base level using fluorescence resonance energy transfer~\cite{Cisse2012,Chen2007,Didenko2001} and fluorescence correlation spectroscopy~\cite{Chen2008, Altan-Bonnet2003, Weiss1999}.

\begin{figure}
    \includegraphics[width=\columnwidth]{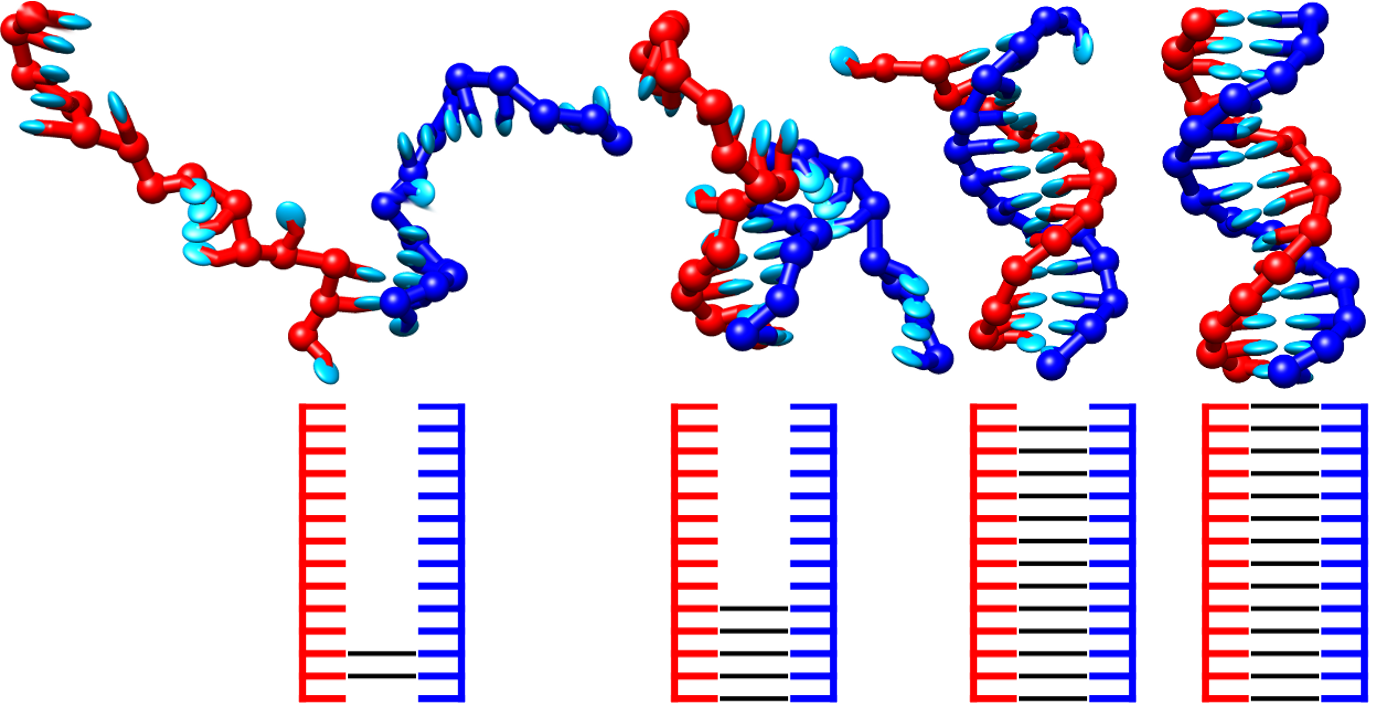}
    \caption{Top: Snapshots from a typical hybridization trajectory of a coarse-grained oxDNA2 simulation.
    Each strand has 14 bases, and each base is visualized by two particles:
    a sphere that represents the backbone and a disc that represents the base.
    The two strands make an initial contact near the end of the strands and then zip up to a fully hybridized state.
    Bottom: Schematic representation of the inter-strand hydrogen bonding pattern. 
    Red and blue lines represent the two strands, and black lines denote hydrogen bonds.
    }
    \label{fig:chimera_example}
\end{figure}

Molecular Dynamics (MD) computer simulations can provide insight into the hybridization mechanism on a base-by-base level. While an atomistic description of the two strands, water, and counterions would provide the highest resolution, it is computationally too expensive to obtain a large dataset of hybridization trajectories. Coarse-grained models sacrifice detail for speed, and several such models of DNA can be found in the literature. In this work we use the recently developed oxDNA 2 model, in which each nucleotide is described by three interaction sites: a backbone site, a hydrogen bonding site, and a stacking site (Figure~\ref{fig:chimera_example} top).
It is parametrized to match the melting temperatures of the SantaLucia model~\cite{SantaLucia2004} at various salt concentrations, as well as physical characteristics such as major-minor grooving and radius of gyration~\cite{Snodin2015}.
Interactions in the oxDNA2 model include stacking, hydrogen bonding and electrostatics.
It has been used to study processes such as hybridization~\cite{Ouldridge2013} and toehold-mediated strand displacement~\cite{Srinivas2013}.
In their study on hybridization, Ouldridge and coworkers used Forward Flux Sampling to guide the system toward the duplexed state~\cite{Ouldridge2013}.
They found a mechanism consistent with a two step process of nucleation and zipping up similar to the classical mechanism.

Here we adopt a different approach to obtaining mechanistic information about the hybridization pathway from coarse-grained simulations. Starting with a large set of hybridization trajectories we construct a Markov State Model (MSM) that describes the stochastic dynamics of the system across a large set of discrete states that represent the two strands' hydrogen bonding pattern. The generation of MSMs from MD simulations to describe the dynamics of complex systems has been pioneered in the context of protein folding~\cite{Bowman2009b,Noe2009,Voelz2010,Pande2010}.
Having obtained an MSM, we use the framework of Transition Path Theory (TPT)~\cite{Metzner2009} to obtain quantitative information about the ensemble of pathways by which DNA hybridizes.

\section*{\label{sec:methods}Methods}

\subsection*{\label{subsec:oxdna}Coarse-Grained Molecular Simulations}

We use the coarse grained oxDNA2 model to simulate DNA hybridization~\cite{Snodin2015}.
In this model, the solvent is treated implicitly, and each nucleotide is represented by three particles: one for the DNA backbone and two for the nucleobase.
It is parametrized to reproduce the duplex melting temperatures of the SantaLucia model~\cite{SantaLucia2004}, which it does accurately~\cite{Snodin2015}.
It incorporates a limited amount of sequence specificity by considering base-dependent stacking interactions. The strength of inter-base hydrogen bonds, on the other hand, is considered to be the same for A-T and G-C base pairs, and mismatched bases cannot form hydrogen bonds.

We simulated two ($\alpha = 1,2$) complementary strands of DNA with $N=14$ bases each.
The sequences are 
\begin{center}
\texttt{3'-GCTGTTCGGTCTAT-5'} \\
\texttt{5'-CGACAAGCCAGATA-3'}
\end{center}
and are designed to limit non-intended base-pairing~\cite{Ouldridge2013}.
At the beginning of a trajectory, the two strands were placed at random positions in a cubic simulation box with 10 nm side length. The dynamics of the system was propagated for up to 3 microseconds or until the strands were fully hybridized, i.e., each base formed a hydrogen bond with its complementary pair. Two bases were considered hydrogen bonded if their hydrogen bond contribution to the interaction energy was larger than 1.79 kcal/mol as done in previous work~\cite{Ouldridge2013}.
The simulation timestep was 15 femtoseconds. Temperature was held fixed at 300 Kelvin using an  Anderson-like  thermostat that stochastically resets a particle's momentum, using a collision time of 103 steps and a collision probability of 0.02 as used in previous work~\cite{Ouldridge2013}. Electrostatic interactions where computed for an effective salt concentration of 0.5 M.

\subsection*{\label{subsec:MSM}Construction of the Markov State Model}
In order to create an MSM, we partitioned the raw molecular dynamics data into individual states at each timestep.
The states are defined based on the hydrogen bonding pattern of the two strands.
The bases in each strand were labelled $i,j=1...N$ from the 3$'$ end to 5$'$ end.
At each timestep, we determined which bases are hydrogen bonded with each other.
The hydrogen bonding pattern only includes bonding between the two strands; instances of intra-strand bonding were ignored.
If a base has hydrogen bonding interactions with more than one other base, only the base with the strongest interaction was considered.
From this list of hydrogen bonds, we labelled each state $s$ with a number $m$ that encodes the hydrogen bond sequence:
\begin{equation}
    m(s) = \sum^{N}_{i=1}(N+1)^{i} \sum^{N}_{j=1}j\,\omega_{1 i j}(s)
    \label{eq:encoding}
\end{equation}
where
\begin{equation}
    \omega_{\alpha i j}(s)=
    \begin{cases}
    1& \text{if base } i \text{ in strand } \alpha \text{ is bound to} \\
     & \text{base } j \text{ in strand } 3-\alpha ;    \\
        0& \text{otherwise.}
    \end{cases}
\end{equation}
This encoding scheme ensures that each state label is unique, and that we can recover the bonding pattern from the number $m$ alone.
The states $s$ are sequentially labelled from 1 to $M$, the total number of visited states. We adopt the convention that $s=1$ refers to the unbound state, whereas $s = 2, \dots, M$ index states with at least one hydrogen bond in arbitrary order.

Because each base can bind to at most one other base in the opposing strand, the total number of states encoded by~\eqref{eq:encoding} is 
\begin{equation}
    \sum_{k=0}^{N}\ffrac{(N!)^2}{k!((N-k)!)^2} ,
\end{equation}
where $N$ is the number of bases in a single strand. For $N=14$ we obtain approximately $\num{1.6e13}$  distinct states. In practice, however, our simulations visit only $M=8942$  of those, which shows that the vast majority of states are not physically relevant.

Because the entire hydrogen bonding pattern between the two strands is a complex descriptor, we define two observables that provide limited but intuitive information about a state. The first is the net binding count $C(s)$, which is defined as the difference between the number of native base pairs (base pairs that also exist in the fully hybridized conformation) and the number of non-native base pairs:
\begin{equation}
    C(s)=\sum^{N}_{i=1}2\omega_{1i(N+1-i)}(s)-\sum^{N}_{j=1}\omega_{1ij}(s) \label{eq:NBC} .
\end{equation}
In principle the value of $C$ ranges from $-N$ to $N$, with $C=N$ for the fully hybridized state.

The second observable, center of hybridization $D(s)$, reports on the positions of bases in strand $\alpha=1$ that are paired with bases in the other strand:
\begin{equation}
    D(s)=\sum^{N}_{i=1}h_{1i}(s)\left({\frac{N+1}{2}-i}\right) ,
\end{equation}
where
\begin{equation}
    h_{\alpha i}(s)=\sum^{N}_{j=1}\omega_{\alpha i j}(s)
\end{equation}
is one if base $i$ is hydrogen bonded to any base on the other strand, and zero otherwise.
States with positive $D$ have more bonding of strand 1 at the 3$'$ end, while those with negative $D$ have more bonding at the 5$'$ end.

Having established the state space of the Markov model, we proceed by converting each oxDNA simulation trajectories into a sequence $m_1, m_2, \dots$ of hydrogen bonding classifiers. The time $\Delta$ between steps in this sequence is determined by the output frequency of the coarse-grained simulations. We then use the MSMBuilder software package~\cite{Beauchamp2011} to construct from these sequences a family of MSMs, parametrized by the lag time $\tau$ (which must be a multiple of $\Delta$). This is done in two steps. First we determine the number of transitions between states that are a time $\tau$ apart using a sliding-window counting scheme. From this count matrix we obtain the transition probability matrix $T$ using a Maximum Likelihood Estimator that enforces detailed balance~\cite{Prinz2011}.

For each MSM we then solve the eigenvalue equation
\begin{equation}
    \psi_{i} T = \lambda_{i}\psi_{i} \label{eq:eigenequation},
\end{equation}
i.e., $\psi_{i}$ is $i$-th left eigenvector of $T$ and $\lambda_{i}$ is its eigenvalue.
If $T$ is a regular stochastic matrix, then the Perron-Frobenius theorem guarantees that the largest eigenvalue (which we take to be $\lambda_1$) is equal to one, and the corresponding eigenvector $\psi_1 \equiv \pi$, when properly normalized, is the equilibrium probability distribution. All other eigenvalues are less than one in magnitude, and the corresponding eigenvectors describe dynamical processes that redistribute probability among the states. These processes relax on the so-called implied time scale
\begin{equation}
    t_i = -\frac{\tau}{\ln \lambda_i}
    \label{eq:eigenvalue_time} .
\end{equation}

The parameter $\tau$ that enters the construction of an MSM must be carefully chosen. One typically desires a small lag time so that the MSM has the highest possible temporal resolution. However, if $\tau$ is too short then the MSM does not accurately describe the dynamics of the original system. 
To find the shortest acceptable lag time we use a basic property of MSMs~\cite{Swope2004}: the time scales~\eqref{eq:eigenvalue_time} are independent of $\tau$ if the latter exceeds the time over which the system dynamics becomes Markovian. We therefore build MSMs for a range of lag times, and calculate their slowest relaxation time scales. We then choose the lag time as the time at which the implied time scales become independent of $\tau$.

\subsection*{\label{subsec:TPT}Transition Path Theory}

Once an accurate MSM has been built, one can use different approaches to analyze its behavior in addition to the eigenvector decomposition~\eqref{eq:eigenequation}. We use Transition Path Theory~\cite{Noe2009,E2006,Metzner2009} to obtain statistical information about the DNA hybridization mechanism. 
A transition path connects a set of reactant states, $A$, to a set of product states, $B$, which we take to be the fully unbound and the fully hybridized state, respectively. All other states form the set of intermediates, $I$.
Using the TPT functionality of the MSMBuilder software~\cite{Beauchamp2011}, we calculate for each state the forward committor, $q_i^+$, which is the probability of reaching state $A$ before reaching state $B$ when starting from state $i$. Reactant and product states have committor values of 0 and 1, respectively, and the committor for the remaining states can be obtained by solving the set of equations~\cite{Noe2009}
\begin{equation}
    -q_i^++\sum_{j\in I}T_{ij}q_j^+=-\sum_{j\in B}T_{ij}
\end{equation}
The forward committor is a measure of how far along a transition any individual state is, and is therefore the ideal reaction coordinate. An analogous quantity, the backward committor, is the reaction coordinate for the inverse path. For a transition matrix that obeys detailed balance, the backwards committor is $q_i^-=1-q_i^+$.

TPT allows us to calculate which pathways contribute most to DNA hybridization by considering the flux between two states, which is the expected number of transitions between those states within one timestep.
Limiting ourselves to reactive paths that connect $A$ and $B$, the flux between two states $i$ and $j$ is~\cite{Noe2009}
\begin{equation}
	 f_{ij} = \pi_i q^-_i T_{ij} q^+_j
\end{equation}
where $\pi_i$ is the equilibrium population of state $i$.
Trajectories can go back and forth between two states, which builds up flux between those states but this built-up flux does not contribute towards reaching the product states $B$.
The net flux~\cite{Noe2009}
\begin{equation}
    f_{ij}^+=\mathrm{max}(f_{ij}-f_{ji},0)
\end{equation}
removes the effects of such recrossings. The capacity of an entire path is defined as the smallest net flux of each of its transitions. With these quantities at hand, one can calculate the highest capacity pathways that connect reactive and product states.

A related property is the ``fraction visited'' of a state, which is the fraction of reactive paths that visit the state along the way~\cite{Dickson2012}. This quantity can also be calculated using the MSMBuilder software.

One way we calculate rate information is through the mean first passage time (MFPT). We compute the MFPT to reach the fully hybridized state for all other states by solving the equation~\cite{Singhal2004}
\begin{equation}
    (E-T')\vek{b}=\vek{c}
    \label{eq:mfpt}
\end{equation}
where $T'$ is a matrix with elements
\begin{equation}
    T'_{ij}=
    \begin{cases}
        2, & \text{if } i \in B \text { and } j=i\\
        0, & \text{if } i \in B \text{ and } j \ne i\\
        T_{ij}, & \text{otherwise,}
    \end{cases}
    \label{eq:T'}
\end{equation}
$E$ is the MxM identity matrix,  $\vek{b}$ is the vector of mean first passage times in units of the lag time $\tau$, and $\vek{c}$ is an M-length column vector containing all ones except at the sink states, where it is 0. 

As an alternative to the MFPT, we also calculate the time scale $t_{A\rightarrow B}$ of the hybridization process, defined as~\cite{Noe2009}
\begin{equation}
    t_{A\rightarrow B}=\ffrac{ \sum_{i=1}^{M} \pi_i q_i^-}{\sum_{j=2}^{M}\pi_1 T_{1j}q_j^+} \tau
    \label{eq:flux_time} .
\end{equation}
Here, the denominator corresponds to the expected number of transitions between $A$ and $B$ per time $\tau$ in either direction, and the numerator is the fraction of paths in the forward direction.

\section*{\label{sec:results}Results}

\begin{figure}
    \includegraphics[width=\columnwidth]{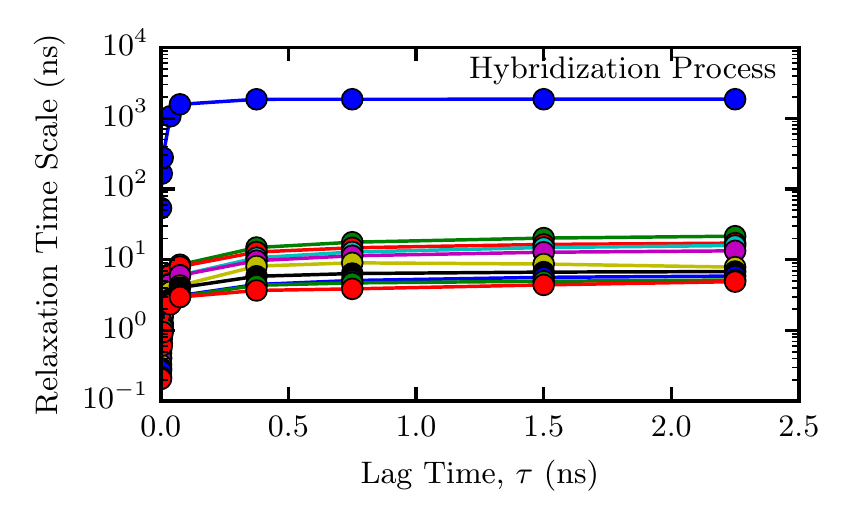}
    \caption{The ten longest relaxation time scales of MSMs built with increasing lag time $\tau$.
	Only at sufficiently long lag times the relaxation time scales become independent of $\tau$, which is a necessary condition for the system dynamics to be Markovian. The process with the longest relaxation time scale corresponds to the transition from the fully unbound to a hybridized state (Table~\ref{tab:first_eigenvector}).
    }
    \label{fig:implied_timescales}
\end{figure}

\begin{table}
    \begin{tabular}{>{\centering}m{2cm}|>{\centering\arraybackslash}m{2cm}}
    State & Contribution to Slowest Process
    \\\hline
        \includegraphics[width=2cm,height=1cm]{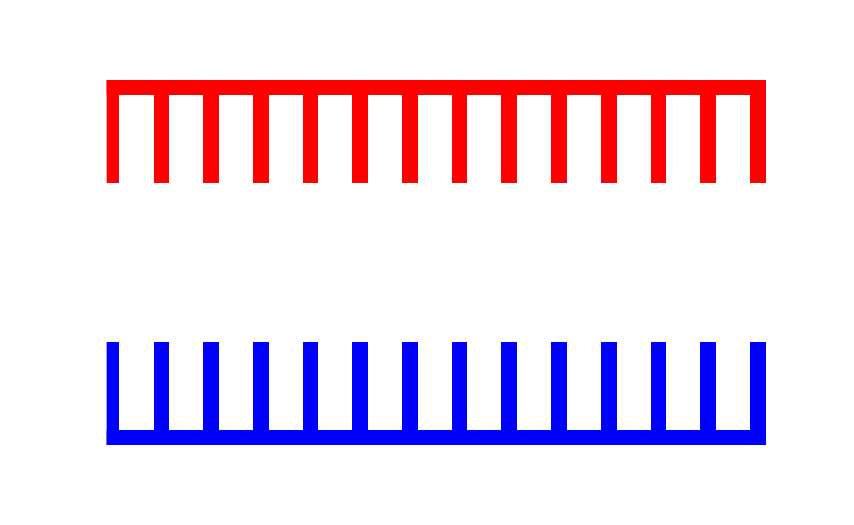} & -1.00 \\\hline
    \includegraphics[width=2cm,height=1cm]{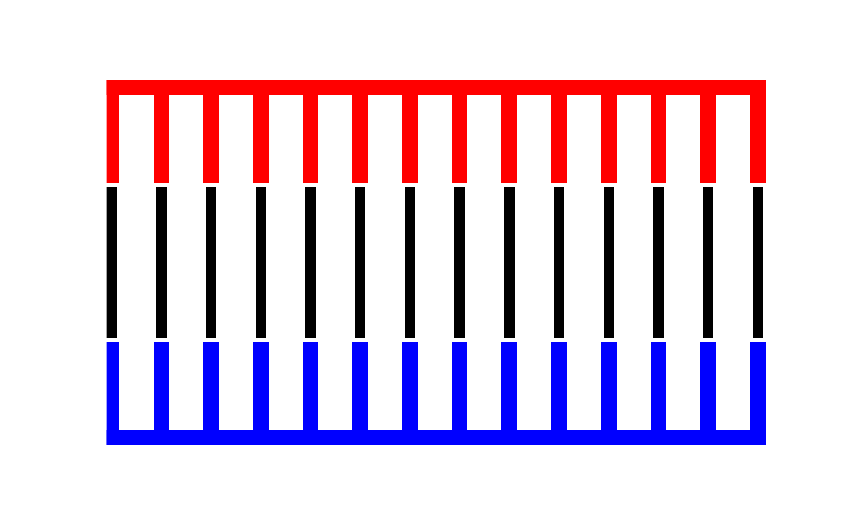} & 0.46 \\\hline
    \includegraphics[width=2cm,height=1cm]{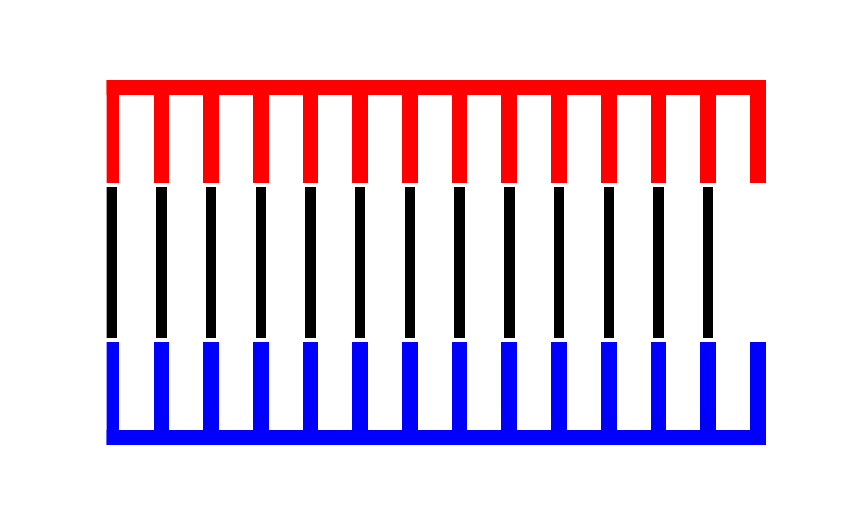} & 0.16 \\\hline
    \includegraphics[width=2cm,height=1cm]{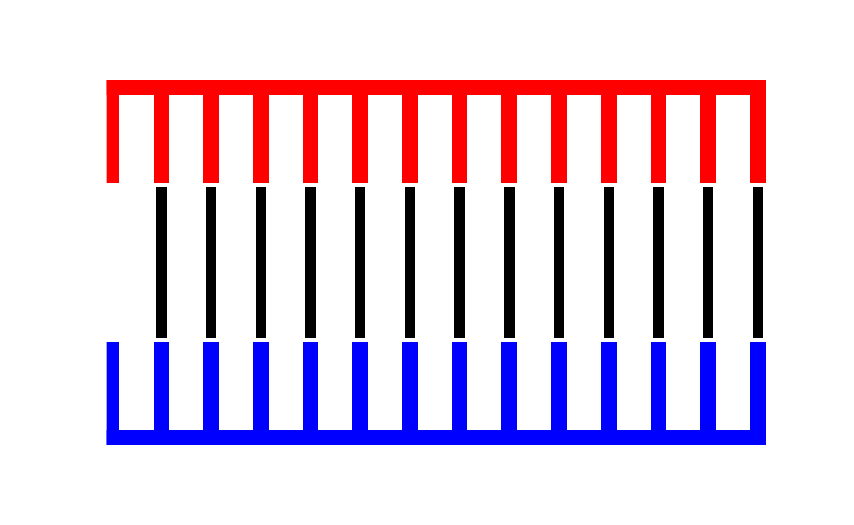} & 0.16 \\\hline
    \includegraphics[width=2cm,height=1cm]{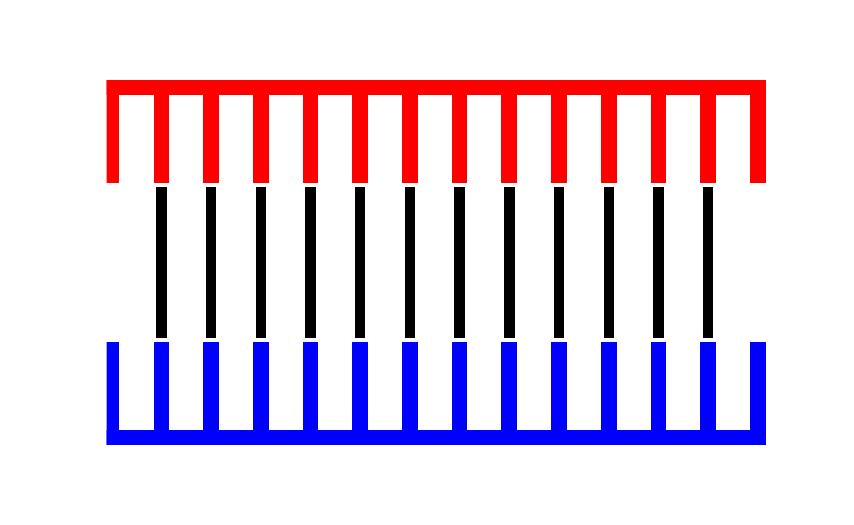} & 0.05 \\\hline
\end{tabular}
\caption{
Five states with the largest (by magnitude) contributions to the eigenvector of the process with the longest relaxation time scale. The eigenvector is normalized so that the unbound state has a contribution of -1. This process mostly shifts probability from the unbound state to the completely hybridized and three mostly hybridized states. 
}
\label{tab:first_eigenvector}
\end{table}

We begin with a large number of hybridization trajectories obtained using the oxDNA model. Using the mapping~\eqref{eq:encoding}, we convert them to  trajectories in the discretized state space that describes the two DNA strands' hydrogen bonding pattern. As described in the Methods section, we construct a series of MSMs from these discretized trajectories by counting transitions that occur over the lag time $\tau$. If the Markovian dynamics described by an MSM is an accurate approximation of the underlying oxDNA dynamics, then the relaxation time scales~\eqref{eq:eigenvalue_time} are independent of the lag time. To find the smallest value of $\tau$ for which this is the case, we plot in Figure~\ref{fig:implied_timescales} the implied time scales for the ten slowest relaxation processes of MSMs constructed using different lag times. We find that $\tau = \SI{0.75}{ns}$ is the shortest acceptable lag time, and use the corresponding MSM in the following analysis.

Figure~\ref{fig:implied_timescales} also shows that there is a large, 100-fold difference in the time scale between the slowest relaxation process and the second-slowest one. To illustrate the nature of this lengthy process, we show in Table~\ref{tab:first_eigenvector} the five states that have the largest contribution to the corresponding eigenvector. The eigenvector is defined only up to a multiplicative constant, which we chose such that the unbound state has a contribution of -1. We see that the slowest process describes transformations between the unbound state on one side and completely or mostly hybridized states on the other. Among the latter, the perfect duplex has the largest weight, but states that show fraying at either or both ends also contribute significantly to this process.

\begin{figure}
    \includegraphics[width=\columnwidth]{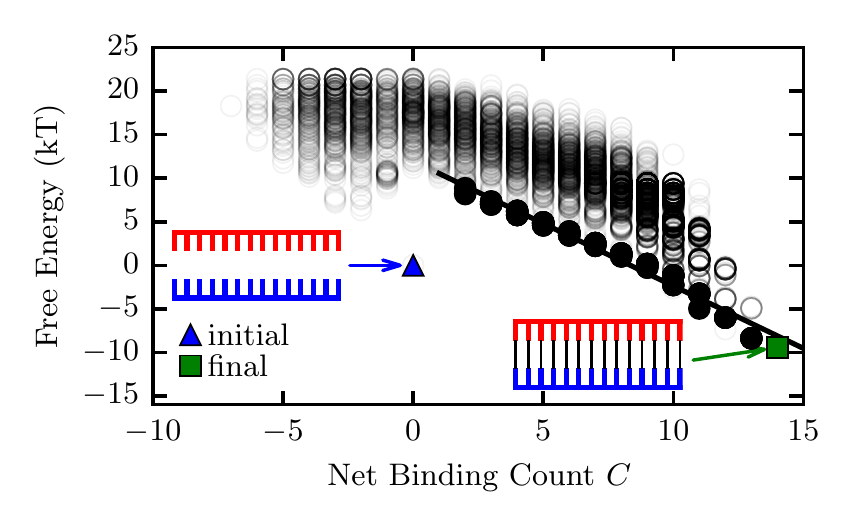}
    \caption{Free energy of each state, projected onto the Net Binding Count $C$.
        The free energy was calculated from the equilibrium probability distribution.
        Areas with more states appear darker.
		States with a large number of bound bases tend to have a low free energy.
    The filled points are the states that appear in the 20 highest capacity paths.
    The solid black line is the line of best fit of the filled points.
}
    \label{fig:free_energy_vs_binding_count}
\end{figure}

The eigenvector of the transition matrix that has an eigenvalue of 1, when properly normalized, is the vector of equilibrium probabilities, $\pi$. From it, we compute the difference in free energy between the unbound and the $i$-th state as
\begin{equation}
    \Delta G_i=-\kT\,\mathrm{ln}\left(\ffrac{\pi_i}{\pi_1}\right) .
\end{equation}
Figure~\ref{fig:free_energy_vs_binding_count} shows these free energies projected onto the net binding count $C$, defined in equation~\eqref{eq:NBC}. We find that the initial, unbound state has a significantly lower free energy than any other state with a net binding count close to zero. The reason for that is the loss of translational entropy when going from two independently diffusing strands to a single, partially bound duplex. Its magnitude depends on the size of the simulation box, and therefore on the concentration of DNA strands~\cite{Ouldridge2013}. There is a large ensemble of duplex states that vary significantly in free energy, but there is a clear trend of decreasing free energy as the the net binding count increases. There is only a single state with $C=N$, the final and fully hybridized state, which is also the global free energy minimum of the system.

From this graph we estimate the average free energy gain of forming a base pair by calculating a linear fit to the free energies of those states that appear in the twenty paths with the largest capacity (see below). We obtain a negative slope of approximately $1.4 \kT$, or \SI{0.83}{kcal/mol}, per base pair. This is smaller than experimental measurements of base pair formation free energies~\cite{Turner1987,Martin1985,Aboul-ela1985,Freier1986,Kawase1986,Gaffney1984} and oxDNA results obtained at the same temperature~\cite{Ouldridge2013}, but matches oxDNA calculations at elevated temperatures~\cite{Ouldridge2011,Ouldridge2013,Snodin2015}.

In addition to equilibrium free energies, we obtain kinetic information from the MSM. There are multiple ways to measure the time scale of hybridization. First, we use the implied time scale~\eqref{eq:eigenvalue_time} of the slowest relaxation process, which we have shown to correspond to DNA hybridization (Table~\ref{tab:first_eigenvector}). Second, we calculate the mean first passage time~\eqref{eq:mfpt} from the the unbound to the fully bound state. Third, we compute the hybridization time scale from the total flux~\eqref{eq:flux_time}. All three methods yield a consistent estimate for the bimolecular rate constant of $k_{\mathrm{hyb}} = ([\mathrm{DNA}] \, t_{\mathrm{hyb}})^{-1}=\num{1.6e8}$ $\mathrm{M^{-1}s^{-1}}$. This is slower than the rate constant $\num{7.7e8}$ $\mathrm{M^{-1}s^{-1}}$ obtained by Ouldridge and coworkers form oxDNA simulations, but still significantly faster than experimental measurements~\cite{Parkhurst1995}.

\begin{figure}
    \includegraphics[width=\columnwidth]{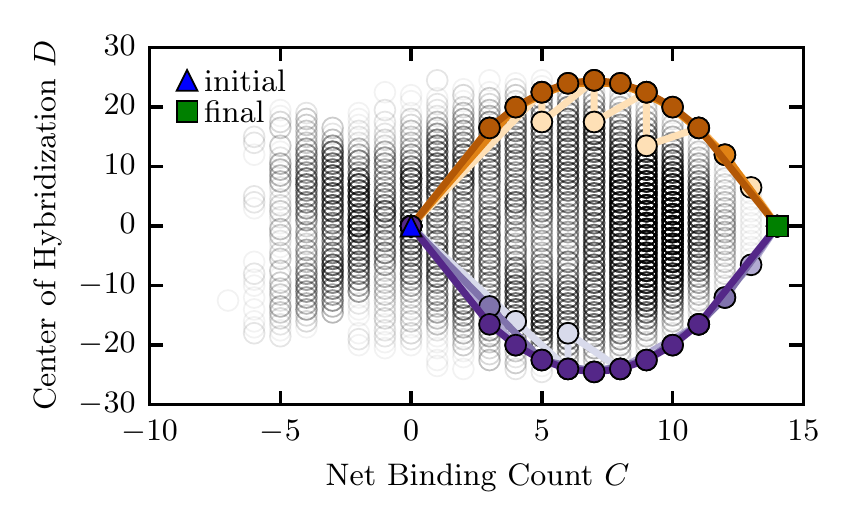}
    \caption{\label{fig:most_likely_paths}Eight paths with the highest flux that connect the unbound state (triangle) to the completely hybridized state (square), shown on a two-dimensional projection of all states onto the net binding count $C$ and the center of hybridization $D$.
    }
\end{figure}

A major advantage of describing the DNA dynamics by a Markov model is that we can obtain statistical information about the hybridization pathways using Transition Path Theory. We use MSMBuilder to calculate the paths with the highest net flux. The top eight paths, projected onto the net binding count $C$ and the center of hybridization $D$, are shown in Figure~\ref{fig:most_likely_paths}. In this two-dimensional  representation, the initial, unbound state is located at $(C,D)=(0,0)$, and the final, fully hybridized state lies at $(N,0)$. The eight highest-flux paths connect these two states by a series of transitions that mainly occur along the upper or the lower boundary of the set of available states. Because each step in the Markov chain corresponds to a physical timestep of length $\tau$, it is possible that hydrogen bonds between multiple bases form or break in a single step.

The location of these highest flux paths reveals the principal mechanism of hybridization: bases initially pair near either end of the strands, and the remaining bases then form hydrogen bonds with their opposing counterparts in a mostly sequential, zipper-like fashion. This process, however, is not entirely rigid: the pathways with the 6th and 7th highest flux contain transitions that appear as vertical lines in Figure~\ref{fig:most_likely_paths}. These transitions correspond to the formation of the next rung in the hydrogen bond ladder, while at the same time the base pair at the end of the strand breaks. Simple end fraying, i.e., the breaking and subsequent reforming of a terminal base pair, occurs frequently in the MSM trajectories but is not apparent in the path analysis, because it does not contribute to the net flux of a path.

\begin{figure}
    \includegraphics[width=\columnwidth]{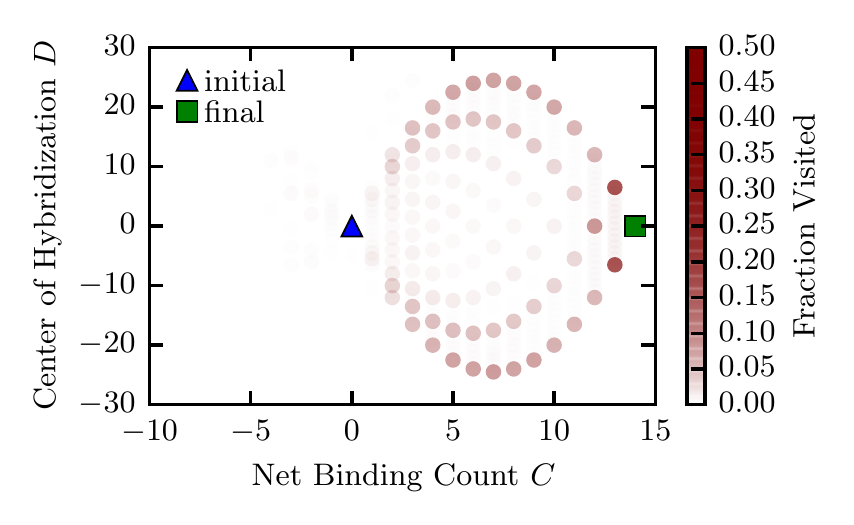}
    \caption{
Illustration of the fraction visited of states along the hybridization pathways. States that are visited more often are shown in darker colors. Multiple nested pathways are discernible, each corresponding to a zipper mechanism. The outer layer contains states with adjacent bases bound, starting from either end of the strand. The next layer contains states in which base pairing starts at the second base from either end.
}
    \label{fig:most_visited_states}
\end{figure}

To identify the states that contribute most to the hybridization process we compute the ``fraction visited'' of each state, defined as the fraction of reactive paths that pass through that state~\cite{Dickson2012}. In the $(C,D)$ representation, these states that are visited most frequently form a series of concentric layers, shown in Figure~\ref{fig:most_visited_states}.  States in the outer layer, which have the highest probability of being visited in a reactive trajectory, contain a single, consecutive segment of hybridized bases that begins on one end of the strands. The hybridization pattern of states in the second layer is similar, except that paired segment starts at the second-to-last base on either end of the strand. These states with a frayed end have a lower probability of being visited. States in even more central layers show more extensive end fraying. This figure illustrates that states with a single block of hybridized base pairs, which can be reached for example by the step-wise base pairing in a zipper-like mechanism, have the highest likelihood to be visited along a reactive trajectory.

Transition Path Theory is built around the committor $q_i$, which is defined as the probability that a trajectory starting from state $i$ will reach the product state $B$ before it reaches the reactant state $A$. It describes the position of a state along the $A \rightarrow B$ reaction, and is considered the ideal reaction coordinate. It is completely determined by the choice of reactant and product states and the rules of the underlying dynamics, and does not require an arbitrary choice of a low-dimensional collective variable to measure the reaction progress. On the other hand, it can be difficult to obtain physical meaning from a set of committor values. 

\begin{figure}
    \includegraphics[width=\columnwidth]{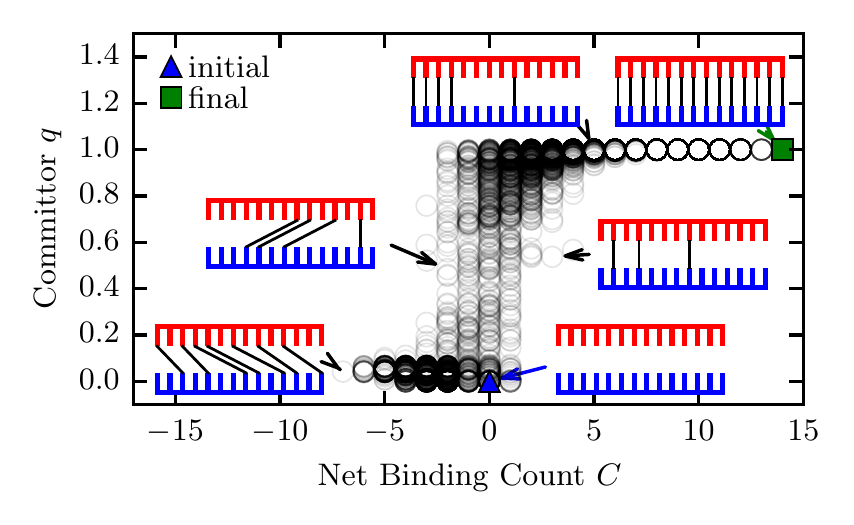}
    \caption{\label{fig:committors}
Committor value of each state, projected onto the net binding count. Higher opacity indicates a larger number of states. The committor, a measure of progress towards the hybridized product state, increases rapidly at small values of $C$. States with the same net binding count can have very different committor values.
}    
\end{figure}

To measure the extent to which the committor values correlate with the physically intuitive net binding count, we show in Figure~\ref{fig:committors} a scatter plot of these two quantities. By construction, the unbound and the fully hybridized states have committor values of 0 and 1, respectively. We see that all states with $C \gtrsim 4$ have a committor value of nearly unity, which means that these states are almost certain to evolve towards a fully hybridized DNA duplex. States with $C \lesssim -4$, on the other hand, will almost always separate into two isolated strands. We find that states with a net binding count close to 0 show a large variability of their committor values, which span the entire range from 0 to 1. In this region, where states have a similar number of native and non-native bonds, it is not possible to infer the committor value from a measurement of the net binding count alone. In particular, we cannot select the subset of states with $q = 1/2$ by specifying $C$ alone. The net binding count is therefore not a useful reaction coordinate, because one cannot describe the set of transition states with it.

\begin{figure}
    \includegraphics[width=\columnwidth]{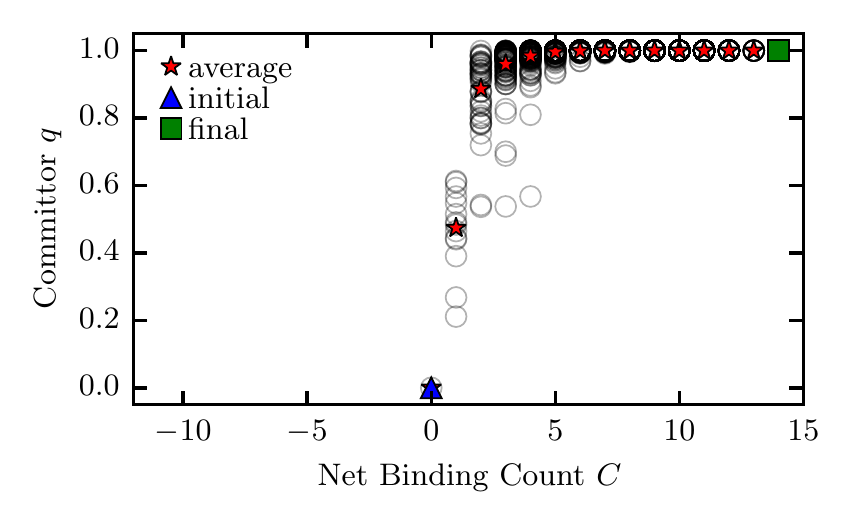}
    \caption{\label{fig:committor_average}
Committor values of the subset of states that have only native base pairs. Considering those states only, the average committor for states with a single, native base pair is 0.5. 
}    
\end{figure}

The predictive power of the net binding count improves if we limit ourselves to only those states that have only native bonds between bases. This is the case for 4805 out of the 8942 states in the MSM. The observable $C$ is then simply the number of such bonds, and its correlation with the committor is shown in Figure~\ref{fig:committor_average}. We find that the range of committor values for a given choice of $C$ is significantly shorter, and the majority of states with $q \approx 1/2$ have a binding count of 1. Nevertheless, there are also transition states with 2, 3, or 4 native base pairs.

\begin{figure}
    \includegraphics[width=\columnwidth]{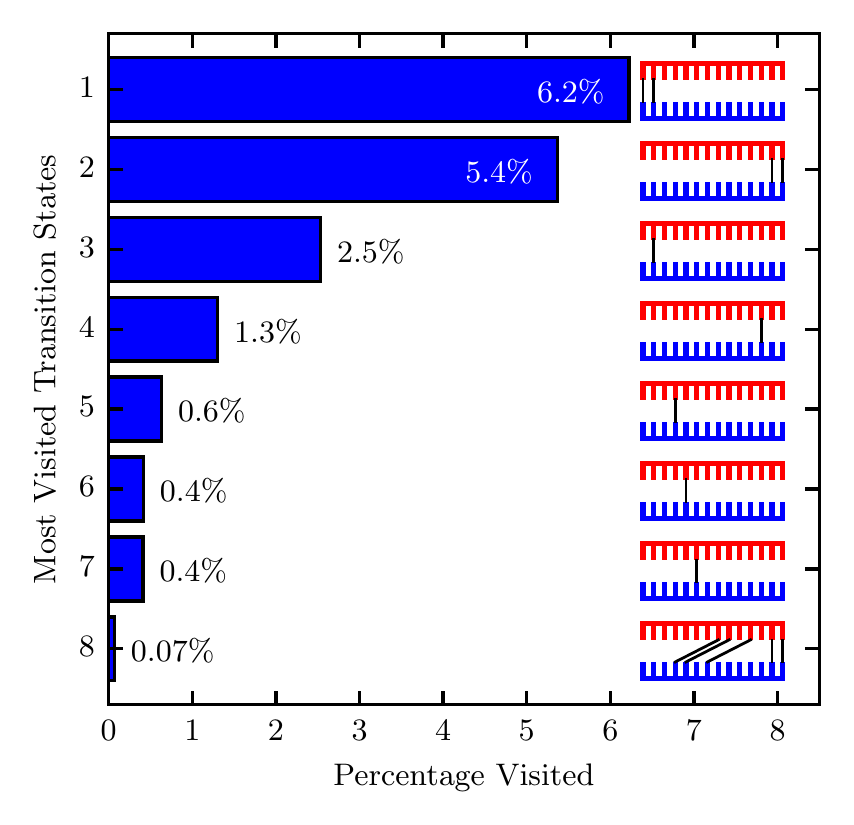}
    \caption{The fraction visited of transition states, i.e., states with committor values between 0.45 and 0.55. Transition states with native base pairs at the ends are often visited along the hybridization pathway.
}
\label{fig:TS_most_visited}
\end{figure}

Returning to the full state space shown in Figure~\ref{fig:committors}, we find that 46 states have a committor between 0.45 and 0.55, which we consider transition states. To identify those that are most important to the hybridization process, we compute the fraction of reactive trajectories that visit these states. Figure~\ref{fig:TS_most_visited} shows the eight states with the highest fraction visited. The top two most visited transition states contain two native base pairs at either end, and together are visited in approximately 11\% of hybridization trajectories. The next five transition states have only a single native base pair at different positions. Following with significantly lower probability, the eighth transition state is the first to contain nonnative in addition to native base pairs.

The sum of these probabilities, i.e., the probability that a hybridization pathway visits any of these states, is significantly less than one. The reason for that is that due to relative large lag time of our model, multiple base pairs can form or break within a single step of the Markov chain. The system can therefore skip these states, and evolve from the unbound to the duplexed state without visiting any transition state along the way.

\begin{figure}
    \includegraphics[width=\columnwidth]{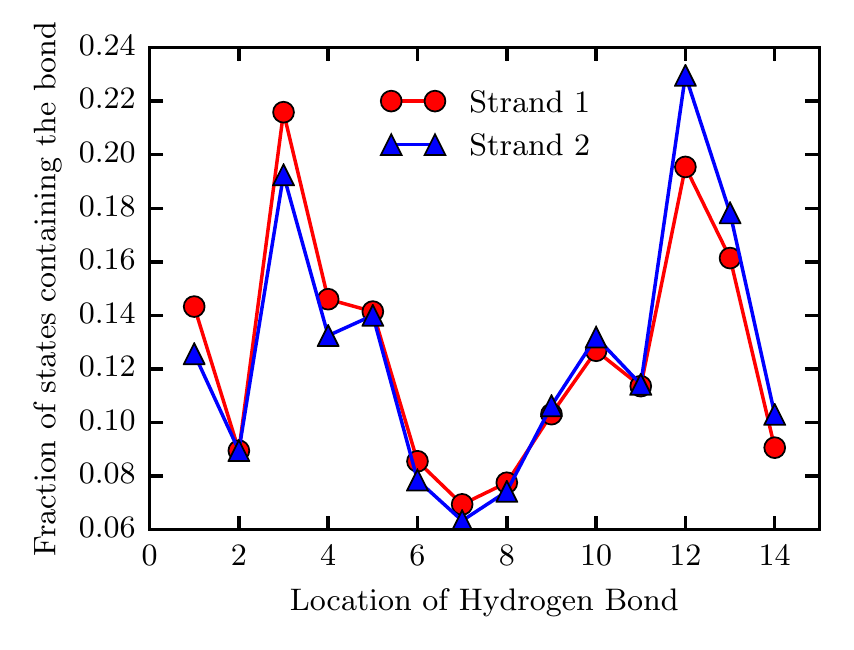}
    \caption{Probability that a base will appear in a first contact of the two strands.
    The third base from either end has a significantly higher probability to participate in the initial base pairing than a base in the middle or at the end of a strand. 
    }
    \label{fig:first_contact}
\end{figure}

The results indicate that the initial formation of the first one or two base pairs is crucial to decide the fate of the newly formed complex. We therefore calculated which bases along the DNA strands are most likely to be involved in those initial contacts. To this end we calculated the probability that base $i$ in strand $\alpha$ is bound to any other base in the ensemble of states that is weighted by their probability to be visited directly from the unbound state:
\begin{equation}
    \ffrac{\sum_{s=2}^{M}T_{1s}h_{\alpha i}(s)}{1-T_{11}}
    \label{eq:first_contact} .
\end{equation}
The results of this calculation are shown in Figure~\ref{fig:first_contact}. We find that the bases at the third position from either end of the strand has the highest probability of being part of the initial base pairing. Bases at the ends or in the middle of the strands have a significantly lower probability. These observations are consistent across both strands.

\begin{figure}
    \includegraphics[width=\columnwidth]{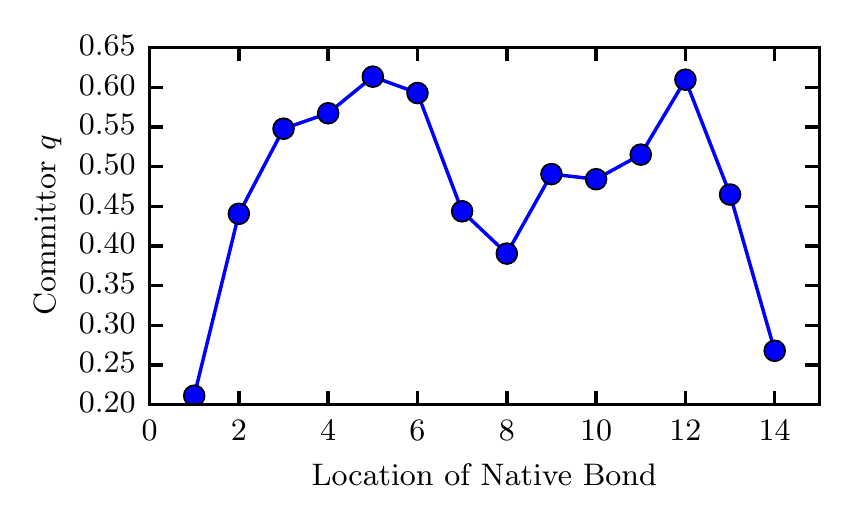}
    \caption{The committor of states that contain only a single native base pair and no non-native base pairs.
 If the single base pair is at the end or in the middle then the committor is low, indicating that this nascent duplex is more likely to fall apart than to proceed to full hybridization.
}
    \label{fig:one_bond}
\end{figure}

Having established that bases close to, but not directly at, the ends of the DNA strands are more likely to participate in the initial base pairing, we now study the effect of these base pairs on the probability of successful hybridization. Figure~\ref{fig:one_bond} shows the committor values of the 14 states that have a single, native base pair. We find that if this base pair is located positions 3--6 or 12, then the probability that the DNA will fully hybridize is large. If the base pair forms in the middle or at either end, then the nascent duplex is more likely to fall apart.

\section*{Discussion}
We investigated how DNA hybridizes and identified the important states and pathways in going from ssDNA to dsDNA.
We identified the important states and pathways by simulating coarse-grained DNA and creating an MSM from the simulation data.
Our results suggest the most important transition states contain one or two native base pairs.
To have a high likelihood of binding, we only required 3 native base pairs.
Our most important pathways also show a zipping up process as suggested by previous work~\cite{Wetmur1968, Porschke1971}.
Our results suggest that once a few bases are bound, DNA is likely to finish binding fully. 
The DNA tended to bind from one end instead of in the middle (Figure \ref{fig:most_visited_states}).
That suggests the identity of the bases on the ends could control the kinetics more than those in the middle.
The number of bases required before full hybridization in our model seems to be smaller than in other experiments~\cite{Porschke1971, Cisse2012}, but is consistent with results in previous oxDNA2 simulations~\cite{Ouldridge2013}.

One strength of our approach was building up a model from simulation data.
From this, we were able to get information on the large number of potential paths and the kinetics associated with those paths.
By the nature of this problem, there is a large state space of theoretically $\num{1.6e13}$ states but we only access 8952 states.
This shows the number of relevant states for hybridization is much smaller than the theoretical number of possible states.

Hybridization has been studied using oxDNA2 but using forward flux sampling, a technique for accelerating dynamics~\cite{Ouldridge2013}.
Our results are consistent with those found by Ouldridge and coworkers~\cite{Ouldridge2013}.
Qualitatively, they find a larger probability of initial binding near the ends of the strands.
Additionally they find the probability of reaching the fully hybridized state is larger for base pairs near the middle of the strand in comparison towards the ends.
They also observe a nucleation and zipping process.

Because the MSM is built from oxDNA simulations it can at best be as accurate as this coarse-grained force field. Any deficiency in the underlying force field will propagate into the MSM. 
Other coarse-grained models of DNA are available, for example the 3SPN force field ~\cite{Sambriski2009,Hinckley2013,Hinckley2014}.
Previously it was found that oxDNA and 3SPN exhibit different hybridization pathways~\cite{Ouldridge2013}. Not surprisingly, our MSM is consistent with previous studies of the oxDNA force field.

Another factor that limits the analysis is the choice of variable of interest.
We partitioned the states by hydrogen bonding and that does not seem to be an appropriate reaction coordinate.
In a good reaction coordinate, each value on the reaction coordinate should have a single value of a committor~\cite{Peters2006}.
This property is not seen in our committor plot (Figure \ref{fig:committors}).
For instance at a Net Binding Count of 0, we have committors that range from 0 to 1.
So while extent of hydrogen bonding is a physical and intuitive reaction coordinate, it is not a good reaction coordinate.
It does seem to be valid for states that only contain native base pairs (Figure \ref{fig:committor_average}).

Creating an MSM has an important advantage over some other enhanced sampling techniques such as Umbrella Sampling~\cite{Torrie1977}, Metadynamics~\cite{Laio2002}, or Forward Flux Sampling~\cite{Allen2009}.
In those methods one must first define a coordinate of interest, often called collective variable (CV), before running the simulation. However, it is often not clear beforehand which CV provides a good description of the process of interest, and one would have to perform another set of simulations if one wanted to consider different CVs. 
In our approach there is no need to specify a proposed CV before collecting simulation data. The MSM itself does not describe the system dynamics in terms of a one-dimensional reaction coordinate. Instead it views the dynamics as a sequence of transition in a arbitrarily complex network of states. Where we did make a choice is in the construction of the MSM when we picked a state decomposition based on the pattern of inter-strand hydrogen bonds. Making a different choice would require the calculation of another MSM transition matrix, but it would not require a new set of molecular mechanics simulations which would be computationally expensive.

\section*{Conclusion}

We have shown that the combination of coarse-grained Brownian dynamics simulations, Markov State Models, and Transition Path Theory forms a powerful tool to study the kinetic properties and dynamical pathways of DNA hybridization. Our results are consistent with the classical zipper model: once an initial nucleus of hybridized base pairs is formed, it grows and neighboring bases zip up sequentially. We find the size of the initial nucleus to lie between one and two base pairs, and that in general the net number of base pairs is not a sufficiently informative observable  to characterize the transition state ensemble.

\section*{Acknowledgments}
This work was facilitated though the use of advanced computational, storage, and networking infrastructure provided by the Hyak supercomputer system at the University of Washington.

\bibliographystyle{apsrev4-1}
\bibliography{paper}

\end{document}